\documentclass{elsart}
\usepackage{natbib}
\usepackage{hyperref}
\usepackage{textcomp}
\usepackage{graphicx}
\usepackage{amssymb}
\begin{document}

\begin{frontmatter}


\title{A method to develop mission critical data processing systems 
for satellite based instruments. The spinning mode case.}
%\thanks[footnote1]{This template can be used for all publications in Advances in Space Research.}


\author{Francesco Lazzarotto\corauthref{cor}}
\corauth[cor]{Francesco Lazzarotto}
\thanks[footnote2]{Additional information regarding the corresponding author}
\ead{francesco.lazzarotto@iasf-roma.inaf.it}
\ead[url]{http://www.linkedin.com/in/francescolazzarotto}
%url can be given like this
\author{Sergio Fabiani, Enrico Costa}
\author{Ettore Del Monte, Giuseppe Di Persio, Immacolata Donnarumma}
\author{Yuri Evangelista, Marco Feroci, Luigi Pacciani,}
\author{ Alda Rubini, Paolo Soffitta.}
\address{INAF IASF Roma, via del Fosso del cavaliere, 100. 00133 Rome - Italy.}

\begin{abstract}
% Text of abstract
Modern satellite based experiments are often very complex real-time systems, 
composed by flight and ground segments, that have challenging resource related 
constraints, in terms of size, weight, power, requirements for real-time response, 
fault tolerance, and specialized input/output hardware-software, and they must be 
certified to high levels of assurance. 
Hardware-software data processing systems have to be responsive to system 
degradation and to changes in the data acquisition modes, and actions have 
to be taken to change the organization of the mission operations. 
A big research \& develop effort in a team composed by scientists and technologists 
can lead to produce software systems able to optimize the hardware to reach 
very high levels of performance or to pull degraded hardware to maintain 
satisfactory features. 
We'll show real-life examples describing a system, processing the data of a 
X-Ray detector on satellite-based mission in spinning mode. 
\end{abstract}
\clearpage

\end{frontmatter}

\parindent=0.5 cm

% main text
\section{Introduction}
The introduction was set equal to the abstract

%\section{An equation}
% \begin{equation}
%    \label{eq:1}
% {%
%    \sum_{i=0}^{\infty}A^n\int \mathrm{d}x\, \frac{F_n(x)}{A_n + B_n} =
%    B^n C^n \int\mathrm{d}x\,\int \mathrm{d}y\,
%    \frac{G_n(x,y)}{\mathcal{A}_n{x} + \mathcal{B}_n{y}}
%    }
%  \overfullrule 5pt
%  \mathindent\linewidth\relax
%  \advance\mathindent-259pt
%  \end{equation}

\section{Description of the Algorithm}
The software algorithm we are describing is named SPInning Pipeline 
(SPIPI) and is based on the concept that measurement data recorded by 
a detection system mounted on a spinning satellite contain data related to 
different portions of the field of view that cyclically return to be pointed. 
For our example the satellite is spinning around his axis with an angular 
velocity of 0.8\textdegree per second. 
The total measurement data set is divided both in a temporary grid and in 
a spatial grid. The spatial grid is composed by slices, of square shape 
each of them with the angular dimension of some degrees in the field of 
view (e.g. 6\textdegree). The data related on a given slice is considered during the 
(short) time intervals where the pointing is relatively stable, and then 
merged into temporary event lists related to each given slice. For every 
eventlist related to a given slice of the grid, the attitude correction 
procedure is applied, and then all the imaging procedure is performed.
\section{Reliability issues and used standards}
We designed and implemented The SPIPI software system to analyse data 
acquired by an instrument in spinning mode: the SuperAGILE instrument (SA) of the 
Italian Space Agency AGILE mission, although the instrument was mainly designed 
and optimized to work in pointing mode, with the nominal Attitude Control System 
(ACS) that could maintain attitude stability better than 0.1 degrees/s (see [5]). The 
AGILE ACS went out of order in autumn 2009 due to a failure of the momentum 
wheel. 
All time related data is represented following the ISO 8601 standard (see [1]), for 
the software requirements specifications we used the format suggested by the IEEE 
Standard 830-199 (see [2]). All the log messages generated by the SPIPI software 
are handled using the syslog standard utilities, this opens a lot of possible 
counteractions to logged events [8]. 

\section{Input/Output, scheduling and Archiviation}
The SASPIN procedures was developed to handle the I/O 
for the SPIPI software and to allow processing to be run 
interactively by scientific users and to be automatically 
scheduled by processing work stations. Specifically is 
possible to run the software in source/field mode, 
run daily/threely/weekly ... dataset integrations, log all the 
events and errors. All processing steps, scheduling events, 
processing status and the final detection infos are saved on 
a mysql database. 

\section{Software Summary}
\begin{itemize}
 \item \textbf{What does the SPIPI analysis software take as input?}\\
Satellite attitude data, satellite ephemerides data, photons event lists; 
 \item \textbf{What the program does?}\\ 
Extracts images of the sky to detect sources, dividing the field of view in a grid 
where the pointing can be acceptable for analysis; 
 \item \textbf{What does the program produce as output?} \\
Tables and maps of the detected sky sources, with their positions (RA, DEC) 
coordinates and flux [$\frac{erg}{cm^2 s }$]. 
\end{itemize}
The main results due to the analysis of SA data products generated by the SPIPI 
software is mainly reported in [1]. A future improvement can be to 
rewrite some procedures in a way to output data lists directly on socket streams 
in order to pipe most of them without using temporary files,
and complete the error control system with more recovery functionalities. 
\clearpage
\section{Conclusions and future development}
The main results due to the analysis of SA data products generated by the SPIPI 
software is mainly reported in [1]. A future improvement can be to 
rewrite some procedures in a way to output data lists directly on socket streams 
in order to pipe most of them without using temporary files, 
and complete the error control system with more recovery functionalities.

\begin{figure}
\label{figure1}
\begin{center}
\includegraphics*[width=16cm]{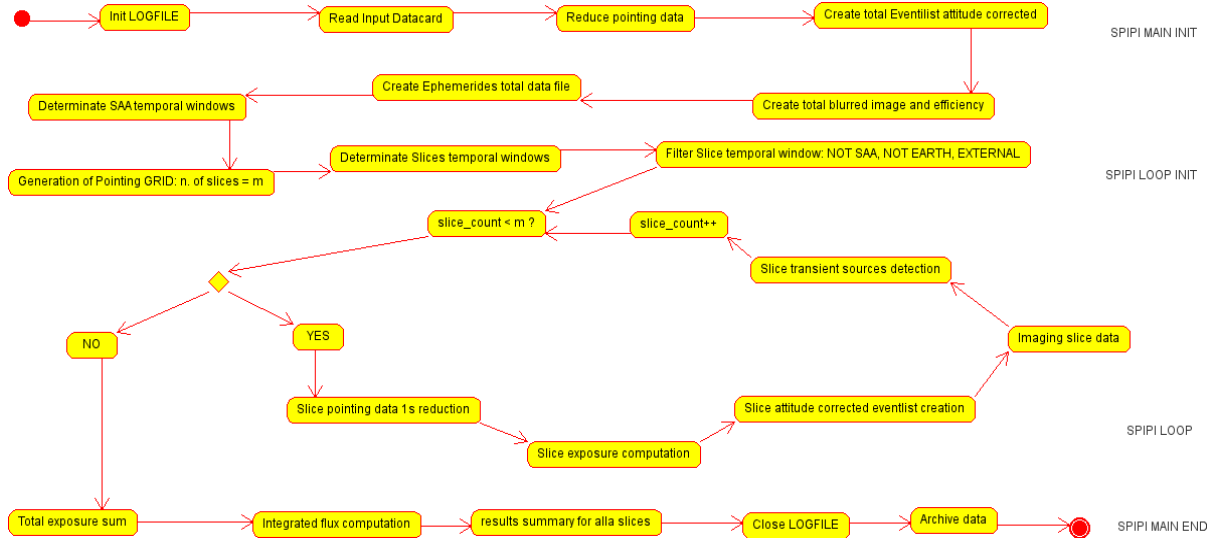}
\end{center}
\caption{SPIPI software activity diagram (flow chart) }
\end{figure}

% \section{Figures}
% 
% As shown in this example, you can provide figures embedded in the text.
% Alternately you can provide all the figures at the end with a separate page
% giving the figure numbers and corresponding captions. Final placement
% of the figures and tables will be done by the typesetters.
% 
% \section{Citations}
% \label{Section 3}
% 
% Remember to give the title of all the articles in references and
% inclusive page numbers in each reference. This is a special
% requirement of ASR. We request authors to provide at least three 
% names before "et al." in multi-authored papers.
% 
% \begin{itemize}
% \item Parenthetical: \verb|\citep{EDM2010}| produces \citep{EDM2010}.
% \item Textual: \verb|\citet{FLSW07}| produces \citet{FLSW07}.
% \item An affix and part of a reference: \verb|\citep[e.g.][Ch. 2]{ISO8601}|   produces \citep[e.g.][Ch. 2]{ISO8601}.
% \end{itemize}

% \appendix

\clearpage

% \begin{table}
% \caption{This is the caption of this table}
% \begin{tabular}{ll}
% \hline
% Parameter&Value\\
% \hline
% Parameter 1 & $526.849 \pm 0.003$ s \\
% Parameter 2 & $5268.49 \pm 0.03$ s \\
% Parameter 3 & $52684.9 \pm 0.3$ s \\
% \hline
% \end{tabular}
% \label{table1}
% \end{table}

\end{document}